\documentclass[a4paper,11pt]{article}
\usepackage{jinstpub}
\usepackage{lineno}
\usepackage{siunitx}
\usepackage[colorinlistoftodos]{todonotes}

\usepackage{defs}
\arxivnumber{2501.05520} % if you have one
\usepackage{acronym}

\makeatletter
\let\@origfpheader\@fpheader
\gdef\@fpheader{\@origfpheader\hfill FERMILAB-PUB-25-0004-CSAID-PPD}
\makeatother

\title{Track reconstruction as a service for collider physics}

\author[a]{Haoran Zhao}
\author[a]{Yuan-Tang Chou}
\author[b]{Yao Yao}
\author[c]{Xiangyang Ju}
\author[d]{Yongbin Feng}
\author[e]{William Patrick McCormack}
\author[a]{Miles Cochran-Branson}
\author[b]{Jan-Frederik Schulte}
\author[b]{Miaoyuan Liu}
\author[f]{Javier Duarte}
\author[e]{Philip Harris}
\author[a]{Shih-Chieh Hsu}
\author[g]{Kevin Pedro}
\author[g]{Nhan Tran}

\affiliation[a]{University of Washington, Seattle}
\affiliation[b]{Purdue University}
\affiliation[c]{Lawrence Berkeley National Laboratory}
\affiliation[d]{Texas Tech University}
\affiliation[e]{Massachusetts Institute of Technology}
\affiliation[f]{University of California, San Diego}
\affiliation[g]{Fermi National Accelerator Laboratory}

\emailAdd{yuan-tang.chou@cern.ch}

\abstract{
Optimizing charged-particle track reconstruction algorithms is crucial for efficient event reconstruction in \ac{LHC} experiments due to their significant computational demands. Existing track reconstruction algorithms have been adapted to run on massively parallel coprocessors, such as \acp{GPU}, to reduce processing time.
Nevertheless, challenges remain in fully harnessing the computational capacity of coprocessors in a scalable and non-disruptive manner.
This paper proposes an inference-as-a-service approach for particle tracking in high energy physics experiments.
To evaluate the efficacy of this approach, two distinct tracking algorithms are tested: Patatrack, a rule-based algorithm, and \exatrkX, a machine learning-based algorithm.
The as-a-service implementations show enhanced \ac{GPU} utilization and can process requests from multiple CPU cores concurrently without increasing per-request latency.
The impact of data transfer is minimal and insignificant compared to running on local coprocessors.
This approach greatly improves the computational efficiency of charged particle tracking, providing a solution to the computing challenges anticipated in the High-Luminosity \ac{LHC} era.
}


\clearpage
\usepackage{aas_macros}
\definecolor{Gray}{gray}{0.9}
\definecolor{Blue}{rgb}{0.88,1,1}
\definecolor{Red}{rgb}{1,0.88,0.88}
\newcolumntype{g}{>{\columncolor{Gray}}p}
\usepackage{blindtext}
\usepackage{graphicx}

\ifcsname ifdothebib\endcsname\else
  \expandafter\let\csname ifdothebib\expandafter\endcsname
                  \csname iffalse\endcsname
\fi

\newboolean{articletitles}
\setboolean{articletitles}{true} % False removes titles in references

\newboolean{uprightparticles}
\setboolean{uprightparticles}{false} %True for upright particle symbols

\newboolean{inbibliography}
\setboolean{inbibliography}{false} %True once you enter the bibliography

\definecolor{mwcolor}{rgb}{0, 0.6, 0}
\definecolor{nsfcolor}{rgb}{0.2, 0.6, 0.6}
\definecolor{cambridgered}{RGB}{163,31,52}
\definecolor{cambridgegray}{rgb}{0.6,0.6,0.6}
\definecolor{cambridgegray2}{rgb}{0.4,0.4,0.4}

\acrodef{AI}{artificial intelligence}
\acrodef{DL}{deep learning}
\acrodef{CNN}{convolutional neural network}
\acrodef{RNN}{recurrent neural network}
\acrodef{GNN}{graph neural network}
\acrodef{GDL}{geometric deep learning methods}
\acrodef{NN}{neural network}
\acrodef{ML}{machine learning}
\acrodef{FML}{Fast Machine Learning}
\acrodef{SM}{standard model}
\acrodef{CMS}{Compact Muon Solenoid}
\acrodef{QCD}{quantum chromodynamics}
\acrodef{LHC}{Large Hadron Collider}
\acrodef{HL-LHC}{High-Luminosity LHC}
\acrodef{QFT}{quantum field theory}
\acrodef{IRC}{infrared and collinear}
\acrodef{EFN}{energy flow network}
\acrodef{EMD}{earth mover's distance}
\acrodef{HPC}{high-performance computing}
\acrodef{A3D3}{Accelerated AI Algorithms for Data-Driven Discovery}
\acrodef{ASAML}{Advancing Science with Accelerated Machine Learning}
\acrodef{HDR}{Harnessing the Data Revolution}
\acrodef{LIGO}{Laser Interferometer Gravitational-Wave Observatory}
\acrodef{MMA}{multi-messenger astrophysics}
\acrodef{HEP}{high energy physics}
\acrodef{HLT}{high-level trigger}
\acrodef{L1T}{level-1 trigger}
\acrodef{BSM}{beyond the SM}
\acrodef{GW}{gravitational wave}
\acrodef{DUNE}{Deep Underground Neutrino Experiment}
\acrodef{LAr}{liquid argon}
\acrodef{TPC}{time-projection chamber}
\acrodef{ZTF}{Zwicky Transient Facility}
\acrodef{FPGA}{field-programmable gate array}
\acrodef{GPU}{graphics processing unit}
\acrodef{IPU}{Intelligence Processing Unit}
\acrodef{TPU}{tensor processing unit}
\acrodef{HLS}{high-level synthesis}
\acrodef{ASIC}{application-specific integrated circuit}
\acrodef{CUDA}{Compute Unified Device Architecture}
\acrodef{SONIC}{Services for Optimized Network Inference on Coprocessors}
\acrodef{L1DS}{level-1 data scouting}
\acrodef{pp}{proton-proton}
\acrodef{URM}{underrepresented minority}
\acrodef{MLP}{multilayer perceptron}
\acrodef{WCC}{weekly-connected-component}
\acrodef{ACTS}{A Common Tracking Software}
\acrodef{L1}{level-1}
\acrodef{MC}{Monte Carlo}

\renewcommand{\paragraph}[1]{\noindent\emph{#1.}}
\begin{document}
\maketitle
\flushbottom

\acresetall

\clearpage
\section{Introduction}
\label{sec:intro}
The computing demands of particle physics experiments at the CERN \acf{LHC}~\cite{Evans:2008zzb}, such as ATLAS~\cite{ATLAS:2008xda} and CMS~\cite{CMS:2008xjf}, are expected to increase dramatically in the era of the \ac{HL-LHC}. This anticipated increase is primarily due to the higher luminosity at the \ac{HL-LHC}, which will lead to more simultaneous proton-proton interactions---known as pileup---in each collision, resulting in a larger number of particles that need to be processed. Consequently, significant efforts are being made to accelerate existing workflows, including event reconstruction, which aims to deduce the properties of particles produced in collisions based on detector measurements.

In collider and fixed target experiments, tracking detectors placed close to the beam collision area and immersed in a strong magnetic field provide high-precision position measurements from which the trajectories of charged particles can be determined.
This task is known as charged particle tracking~\cite{Strandlie:2010zz}.
Among all event reconstruction algorithms, tracking is typically the most time-consuming component, accounting for  45\% or more of the total computing time during data processing~\cite{CERN-LHCC-2020-015,Software:2815292}. By optimizing this critical component, the entire data processing pipeline can achieve faster throughput, enabling more timely analysis and interpretation of the vast amounts of data recorded by the detectors.

Track reconstruction algorithms primarily involve pattern recognition operations to identify the paths of charged particles as they interact with the detector in three-dimensional space. These operations can be significantly accelerated with coprocessors, such as \acp{GPU}, which are well-suited for parallel computing tasks. Integrating these coprocessors with modern CPUs---a method known as heterogeneous computing---enables the system to handle heavy computational loads more efficiently by distributing tasks between CPUs and coprocessors.

To address the challenges encountered when implementing heterogeneous computing for data-intensive physics experiments, an \emph{as-a-service} approach has been developed. The current computing infrastructure aggregates tasks from central experimental operations and individual user requests into a global computing grid, distributing tasks to available computing centers worldwide. However, the traditional technique of directly connecting coprocessors to CPUs in each computing center faces several challenges, such as coprocessor underutilization or overutilization, inconsistent availability of coprocessors across sites, and the need to maintain software compatibility between the varying types of coprocessors and CPUs at each computing site.

\begin{figure}[htp]
\centering
\includegraphics[width=0.8\textwidth]{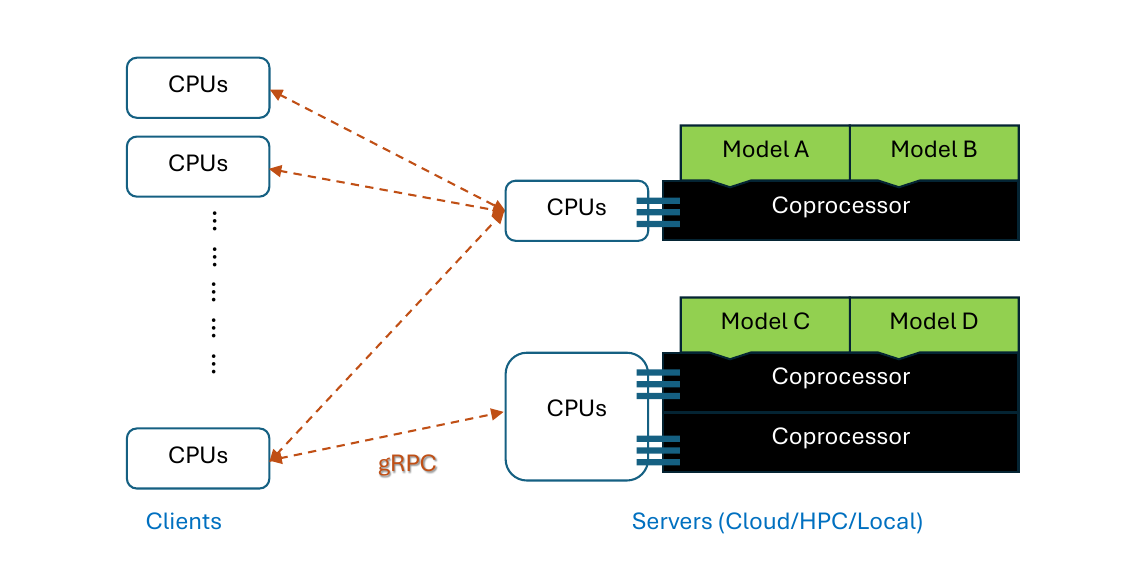}
\caption{Inference \emph{as-a-service} approach: Users send various inference requests from client CPUs, which include details about the type of inference desired, input dimensions and content, and output dimensions and labels. This information is delivered from the clients to the servers through gRPC protocol, a high-performance Remote Procedure Call. The server CPUs receive these tasks, batch them, execute inference on the appropriate coprocessor based on the specific request, and deliver the output back to the client CPUs via gRPC protocol. In this approach, each server can contain a different number of coprocessors and provide different models. Each client can deliver tasks to multiple servers so that the tasks can be processed in parallel. The client-to-server ratio can be scaled based on the demand of client requests.} \label{fig:aas}
\end{figure}

The as-a-service approach overcomes these challenges by adding an abstraction layer that enables the dynamic allocation of CPU and \ac{GPU} resources tailored to specific tasks, as illustrated in Fig.~\ref{fig:aas}. This method provides several key benefits: it helps to achieve optimal \ac{GPU} utilization, facilitates remote access to \ac{GPU} resources, eliminates the need for local GPUs, decouples the server from clients, modularizes software support for CPU and \ac{GPU}. Consequently, server-side coprocessor configuration changes require minimal modification on the client side, simplifying technical support and software compilation processes. This approach has been explored in various technologies including \acp{FPGA}~\cite{Duarte:2019fta, Rankin:2020usv}, \acp{GPU}~\cite{Krupa:2020bwg,Savard:2023wwi}, \acp{IPU}, and CPUs~\cite{CMS:2024twn}, and has undergone extensive testing in numerous experiments such as CMS~\cite{CMS:2024twn}, ProtoDUNE~\cite{Wang:2020fjr,Cai:2023ldc}, and LIGO~\cite{Gunny:2021gne}.

To demonstrate the effectiveness of the as-a-service approach in \ac{LHC} experiments, we focus on two track reconstruction algorithms: Patatrack~\cite{Bocci:2020pmi,Bocci:2020olh}, a rule-based algorithm implemented using \ac{CUDA}, and \exatrkX~\cite{ExaTrkX:2021abe}, a machine learning algorithm based on \ac{GNN}. This paper presents the improvements in throughput achieved for both algorithms by implementing customized server backends that adapt them to the as-a-service model, showcasing a scalable solution to the computational challenges anticipated in the \ac{HL-LHC} era.

In Section~\ref{sec:background}, we review the typical pipeline of track reconstruction algorithms in particle physics, providing a detailed description of both Patatrack and \exatrkX with particular emphasis on their input and output structures. Sections~\ref{sec:patatrackaas} and~\ref{sec:exatrkXaas} discuss the implementation of the as-a-service approach for each algorithm, related performance metrics, including throughput and \ac{GPU} utilization. Finally, we discuss the differences in implementation and the challenges encountered during deployment.

\section{Background}
\label{sec:background}

\subsection{HEP Computing: online and offline reconstruction}
\label{sec:hep_intro}
In typical high-energy physics experiments like ATLAS and CMS at the \ac{LHC}, the reconstruction processes are divided into online and offline reconstruction. \textbf{Online reconstruction} refers to the trigger system that rapidly reconstructs and filters the data. It aims to identify potentially interesting events to record for further analysis~\cite{ATLAS:2020esi,CMS:2020cmk,CMS:2024aqx,Collaboration:2759072,Zabi:2020gjd}. The trigger chain is divided into multiple levels. The \ac{L1T}, implemented in hardware like \acp{ASIC} and \acp{FPGA},  provides the first rapid decision-making layer. It uses a simplified reconstruction of signals from a subset of detector systems to reduce the data rate from 40 \unit{\MHz} to around 750--1000 \unit{\kHz} by selecting events with high-energy particles or specific decay products~\cite{ATLAS:2020esi,Zabi:2020gjd}.

Following the \ac{L1T}, the \ac{HLT} system, which operates in software using commercial CPU or \ac{GPU}s, applies a more refined selection to reduce the event rate to 5--10 \unit{\kHz}. It uses more detailed information and complex algorithms, similar to offline reconstruction, to filter the surviving events by applying additional criteria. Latency and throughput are typically the limiting factors at this stage.

After the online selection, events are written to permanent storage, where \textbf{offline reconstruction} occurs. This stage involves using more sophisticated algorithms to fully reconstruct and calibrate the recorded events. Unlike online reconstruction, offline reconstruction is not constrained by real-time requirements, allowing for the application of precise calibration, alignment corrections, and detailed analysis procedures. This includes fitting charged tracks, identifying jets, and accurately reconstructing decay vertices, etc. Nevertheless, the challenge remains in how to process the vast amount of data efficiently and how to seamlessly integrate new coprocessors into the production framework as the hardware landscape rapidly evolves.

\subsection{Track reconstruction}
When a charged particle passes through a
tracking detector, it leaves behind charge deposits in each detector layer as it traverses the materials. The resulting signals are read out by the detector electronics and then, if initially analog, converted to digital. The primary goal of track reconstruction algorithms is to determine the trajectories of charged particles, including their curvature and point of origin (vertex). The magnetic field applied within the detector causes the trajectory of a charged particle to follow a curved path, with the curvature inversely related to the particle's momentum; tighter curves indicate lower momentum. Accurate vertex determination is crucial for identifying the specific collision event that produced the track, especially in environments with high pileup, such as the \ac{HL-LHC}.

Typical tracking algorithms involve several stages: spacepoint formation, track seeding, track following, and track fitting. Initially, 2D or 3D measurement positions, known as \textit{spacepoints} or hits, are estimated by combining nearby raw detector measurements into clusters and determining their coordinates. Track seeding then uses these spacepoints to form initial track candidates, providing preliminary estimates of trajectory parameters such as direction, origin, and curvature. Track following refines these seeds by adding more spacepoints along the projected path, ultimately leading to the track fitting stage. Here, an initial trajectory is calculated through the spacepoints, enabling the estimation of the particle’s physical and kinematic properties, including charge, momentum, and origin. However, most traditional fitting algorithms, such as the Kalman filter, are generally not optimized to run on \ac{GPU}, limiting how they can scale to handle increasing computational demands.

With the rapid development of \ac{ML}, another class of algorithms using \ac{GDL} has emerged~\cite{Bronstein_2017}. Several architectures exist to utilize 3D point clouds for pattern recognition~\cite{Dezoort:2021kfk,Lieret:2023aqg, Liu:2023siw, ExaTrkX:2021abe,miao2024localitysensitive}. These \ac{ML}-based algorithms learn to cluster or connect spacepoints from the same tracks based on high-dimensional latent space features, leading to improvements in reconstruction speed and accuracy in complex environments with high particle multiplicity.
These algorithms are also more parallelizable by nature.

We demonstrate the feasibility of \textit{tracking as a service} with Patatrack, a \ac{GPU}-optimized rule-based approach described in Section~\ref{sec:patatrackalgo} and the \exatrkX pipeline, a \ac{ML}-based pattern recognition algorithm detailed in Section~\ref{sec:exatrkxalgo}.

\subsubsection{Patatrack}
\label{sec:patatrackalgo}
Patatrack~\cite{Bocci:2020pmi,Bocci:2020olh} is a \ac{GPU}-optimized track reconstruction algorithm designed for the CMS \ac{HLT}~\cite{CMS:2024twn} that employs a cellular automaton for pattern recognition. Unlike traditional tracking algorithms, Patatrack is explicitly tailored for execution on \ac{GPU}, enabling efficient parallel data processing.
The inputs to the algorithm are raw data from the pixel detectors and information about the beam spot---the region where the proton beams overlap during collisions. The algorithm comprises five major sub-algorithms: digitization, hit reconstruction, ntuplet creation, pixel tracks, and vertex reconstruction. The outputs of Patatrack are reconstructed pixel tracks, vertices, and other intermediate objects necessary for downstream tasks such as muon reconstruction.

\begin{figure}[htp]
\centering
\includegraphics[width=1.0\textwidth]{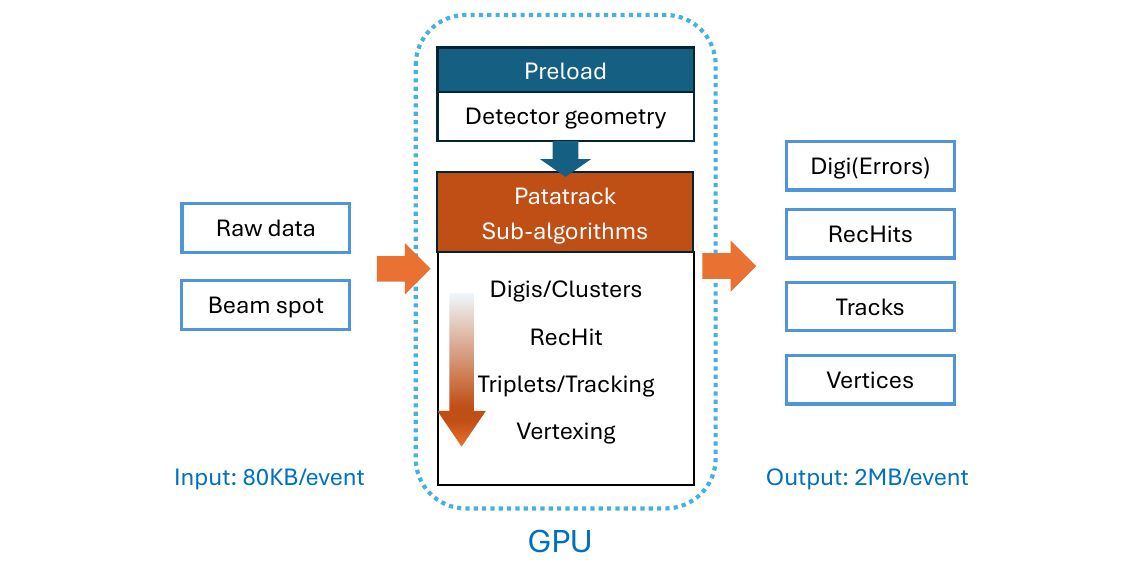}
\caption{This illustration shows Patatrack running on a \ac{GPU} using the as-a-service approach. Common detector-related information, such as the location of each detector layer and detector indices, is preloaded onto the \ac{GPU} to be used later in the processing. For each event, raw data and beam spot information are delivered to the \ac{GPU}. The raw data is compressed before delivery, and during the digitization step, it is unpacked and converted back into ``digis.''
In the same detector layer, neighboring digis are clustered to determine the location of a ``hit,'' representing a single interaction of a charged particle with that detector layer. From the hits, those in adjacent layers are paired to form doublets, then pair to triplets as track seeds. From a seed, a group of hits that potentially form a track are picked, and then a fit is applied to determine the track parameters and vertex location. Finally, the reconstructed tracks and vertices are delivered back to the host.
}\label{fig:patatrackworkflow_2}
\end{figure}

The workflow of the Patatrack algorithm running on a \ac{GPU} with the as-a-service approach is illustrated in Fig.~\ref{fig:patatrackworkflow_2}. To minimize data movement and conversion, intermediate objects are retained on the device, while objects needed for downstream tasks are transferred back to the host. The input size is approximately \qty{80}{\kilo\byte} per event, and the outputs are around \qty{2}{\mega\byte} per event. 

\subsubsection{Exa.TrkX pipeline}
\label{sec:exatrkxalgo}
The \exatrkX algorithm is based on deep learning models. It uses spacepoints as inputs and produces track candidates as outputs. Each track candidate is a list of spacepoints. The pipeline contains three major modules: graph construction, edge classification, and graph segmentation. The graph construction module uses \acp{MLP} to encode spacepoint raw features to a new latent space. The hits from the same charged particles are close to each other and away from other hits from different charged particles. This step is called \textit{embedding}. Then, a fixed-radius algorithm builds connections between spacepoints in this latent space (\textit{edges}). The number of edges after the embedding step necessitates another \ac{MLP} to filter out clearly fake edges---called \textit{filtering}. A fixed threshold is applied to these filtered edge scores to retain true edges while eliminating false ones. The edge classification step uses the interaction network~\cite{gnn:IN}. In the end, the \ac{GNN} edge scores are passed to the \ac{WCC} algorithm to form track candidates.

The \exatrkX pipeline is implemented in \textsc{C++} for CPU- and \ac{GPU}-only~\footnote{The software can be found at https://github.com/The-ExaTrkX-Project/exatrkx-service.}. Trained \textsc{PyTorch} models are executed via the \textsc{libTorch} library. The nearest neighbor search is performed with the \textsc{FRNN} package~\cite{frnn} for \ac{GPU} and \textsc{FAISS}~\cite{douze2024faiss} for CPUs. The connected component algorithm is from the \textsc{Boost} package, with potential future improvements from the \ac{CUDA}-based version in \textsc{cuGraph}~\cite{cugraph}.

The \exatrkX pipeline has been integrated into the \ac{ACTS} framework~\cite{Ai:2021ghi}---an experiment-independent toolkit for track reconstruction. \ac{ACTS} serves as a test bed for a range of tracking and vertex reconstruction algorithms.
Simulation events are generated using Fatras fast simulation~\cite{Edmonds:2008zz}, which invokes \Pythia 8~\cite{Bierlich:2022pfr} to simulate \ttbar \ac{MC} events with an average number of additional proton-proton interactions within the same or nearby bunch crossings (pileup) of 200, replicating the conditions expected at the \ac{HL-LHC}. The raw measurement from detectors is processed and clustered into spacepoints by the framework and provided as inputs to the \exatrkX pipeline. In the model utilized for this study, each spacepoint is characterized by three features, resulting in approximately 350 thousand points per event, corresponding to a data size of average size of \qty{3.4}{\mega\byte} per event. The output from the \exatrkX pipeline includes track candidates with the average size of \qty{1.4}{\mega\byte} per event.

\subsection{Inference as a service using NVIDIA Triton Inference Server}

The inference as a service approach in this paper is implemented using the NVIDIA Triton Inference Server~\cite{triton}. It is an open-source package built on the open-source gRPC server package that standardizes the deployment and execution of \ac{ML} models across various workloads\cite{grpc}. It natively supports several machine learning frameworks as backends: PyTorch~\cite{pytorch}, TensorFlow~\cite{tensorflow}, TensorRT~\cite{tensorrt}, and ONNX Runtime~\cite{onnx}. A backend is a wrapper around an existing \ac{ML} framework that executes client requests and returns the results to the client via gRPC. The Triton server also supports custom implementations using C/C++ or Python, denoted \textit{``custom backend''} in the following discussion. In this paper, we utilized the flexibility of the custom backend to demonstrate the potential to apply it to any custom workflow.

\subsubsection{Custom backend}
\label{exatrkxaas:CustomBackend}

A custom backend may be necessary depending on the complexity of the algorithm pipeline. For example, a complex ML pipeline may consist of multiple modules that do not efficiently align with the ensemble model architecture provided by Triton.~\cite{Zhao:2024uwh}. Data transfer between models on different \ac{GPU}s introduces significant overheads.  
In this scenario, the backend can be meticulously designed to integrate seamlessly with Triton, utilizing its official API to manage heterogeneous computing tasks efficiently. The custom backend also provides a convenient way to scale traditional rule-based algorithms. It serves as a wrapper function around existing algorithms and provides an easy pathway to scaling them with Triton. The use of a custom backend has a few additional benefits:
\begin{itemize}
\item  \textit{Fine-grained control}:  Custom backends allow developers to optimize performance at a low level to avoid unnecessary type conversion and copying in memory, which is preferable for complex pipelines with many models chained together. 
\item  \textit{Low latency}:  Lower latency can be achieved by bypassing the high-level \ac{ML} framework and implementing custom operations to avoid redundant memory copying, which is crucial for real-time applications.
\item  \textit{Custom logic and operations}: Developers can implement specialized operations or algorithms not supported by standard deep learning frameworks.
\item  \textit{Integration with legacy code}: Custom backends can be built from existing C/C++ codebases to avoid unnecessary rewriting and validation.
\item  \textit{Support for non-standard data types:} Custom backends can be designed to handle unique data types or formats not natively supported by the server.
\end{itemize}

\subsubsection{Model performance measurement}
\label{sec:perfclient}
The performance analyzer, ~\verb|perf_analyzer|, is a performance measurement tool provided by the NVIDIA Triton. It is designed to evaluate the throughput and latency of model instances deployed on the server. This analyzer generates and sends a configurable number of inference requests to the Triton server, allowing developers to simulate different load conditions and assess how their models perform under various scenarios. \verb|perf_analyzer|~provides detailed metrics, including response times, latency, and throughput rates, helping developers identify potential bottlenecks and optimize their deployment for better efficiency. The \verb|perf_analyzer| has been utilized in the following standalone studies, along with full LHC workflows, when specified.

\section{Patatrack as a service}
\label{sec:patatrackaas}
\label{sec:pata-aas}

The Patatrack algorithm is implemented as a custom backend that runs on the NVIDIA Triton Inference Server. As shown in Fig.~\ref{fig:patatrackworkflow_2}, the modules in the dark orange block are contained in the backend and are called during the inference stage. When launching the servers, environment data and configurations, such as the cabling map from the front ends to the detector, gain calibrations, and cluster calibration parameters, are preloaded into the server to be used later during inference. Since these values remain consistent during data-taking, they are configured beforehand rather than sent to the server at inference time.

\begin{figure}[ht]
\centering
\includegraphics[width=0.8\textwidth]{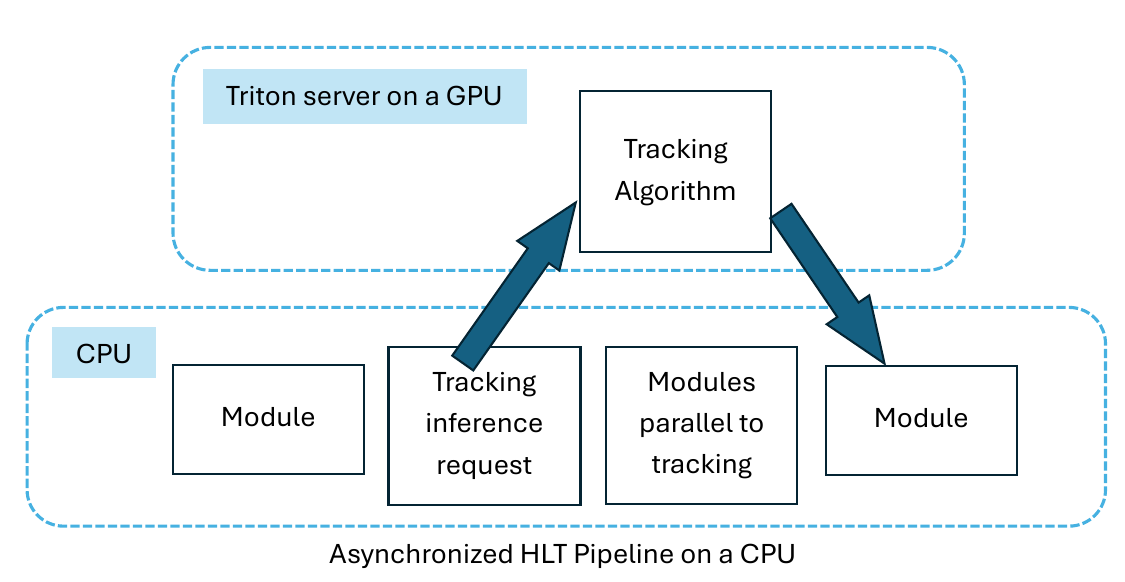}
\caption{Illustration of the \ac{HLT} running with tracking as a service. In the \ac{HLT} workflow, the tracking algorithm inference runs on a server, while other modules in the workflow run on the client CPU in parallel to the tracking algorithms, ensuring no time is wasted waiting for track reconstruction results. This asynchronous pipeline has been implemented in production.}
\end{figure}

\subsection{Standalone algorithm throughput tests}
The throughput of running the algorithm as a service is tested and compared to running the algorithm directly on \ac{GPU}s. In the as-a-service setup, the inference is triggered using dummy client inference calls. The results show that the throughput is found to be 400 events/s with a single-threaded client and 820 events/s when 10 threads are used to communicate with the server. For both tests, the server used is a NVIDIA T4 GPU. The resulting output data rate is found to be approximately \qty{1}{\giga\byte}/s. For a single-threaded client and a direct connect setup, no significant difference in throughput is found when comparing direct GPU with inference as a service. For comparison, the CPU-only rate of Patatrack is found to be 25 events/s on a single 4-threaded \ac{HLT} job.

\subsection{HLT workflow throughput scanning}
\begin{figure}[ht]
\centering
\includegraphics[width=0.7\textwidth]{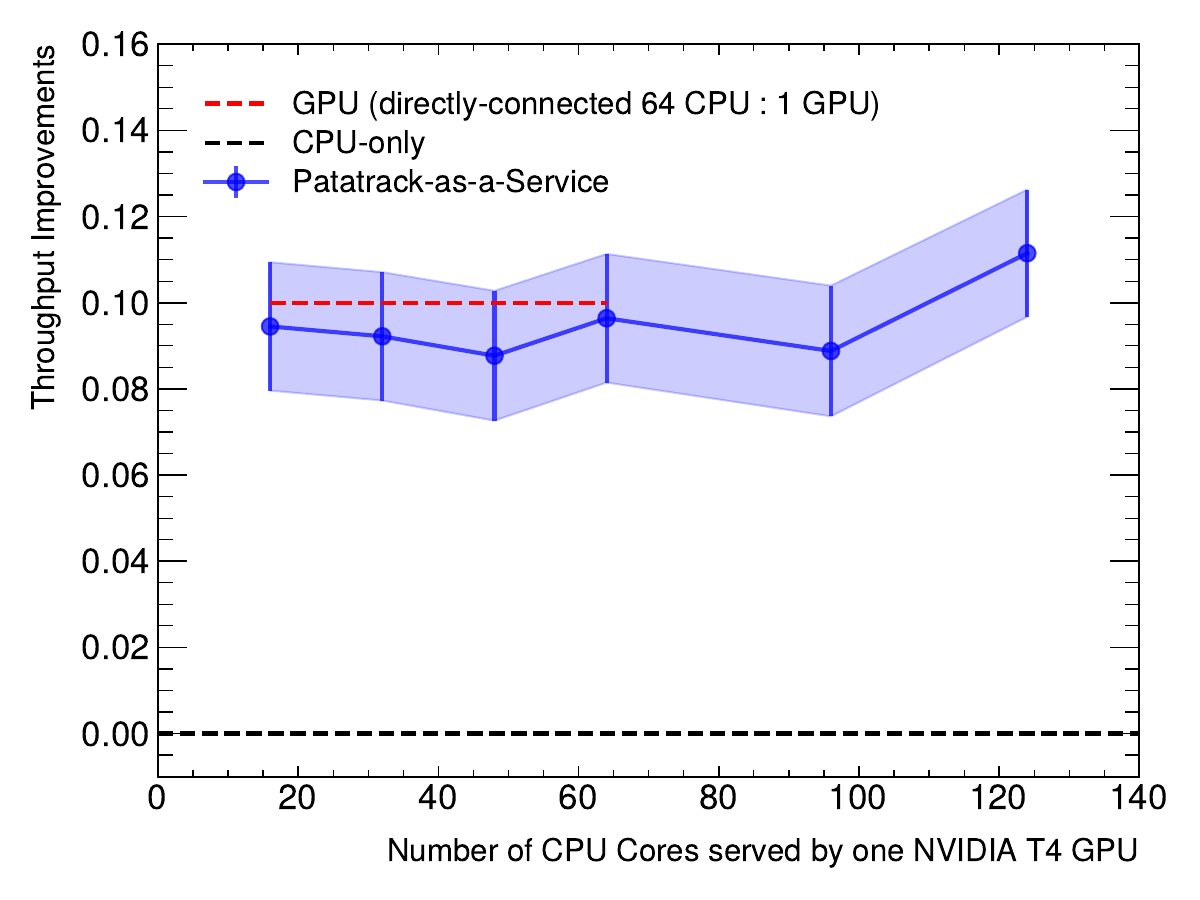}
\caption{A scan of throughput improvement for the \ac{HLT} workflow, varying the number of CPU clients communicating with a single \ac{GPU} server. Direct inference on a \ac{GPU} is limited to 64 CPUs requesting Patatrack inference. This results in about a 10\% throughput gain compared to running it on the CPU alone (black dotted line). Using Patatrack as a service, the system can handle in excess of 120 CPU cores while maintaining the same level of throughput improvement, showing no \ac{GPU} saturation.
}
\label{fig:throughput_scan_patatrack}
\end{figure}

A GPU is considered saturated when increasing the number of CPU clients interacting with the \ac{GPU} server no longer results in increased throughput. To measure the number of CPU clients that can interact with a \ac{GPU} before it becomes saturated, a server with an NVIDIA Tesla T4 \ac{GPU} is used, equipped with the Patatrack backend for inference tasks. On the client side, the CMS \ac{HLT} workflow processes a prepared dataset to simulate actual collisions, with each job using one CPU thread per physical CPU core. The number of CPU clients is increased from 20 to 120, and the resulting throughput improvement is shown in Fig.~\ref{fig:throughput_scan_patatrack}.

\begin{figure}[ht]
\centering
\includegraphics[width=0.7\textwidth]{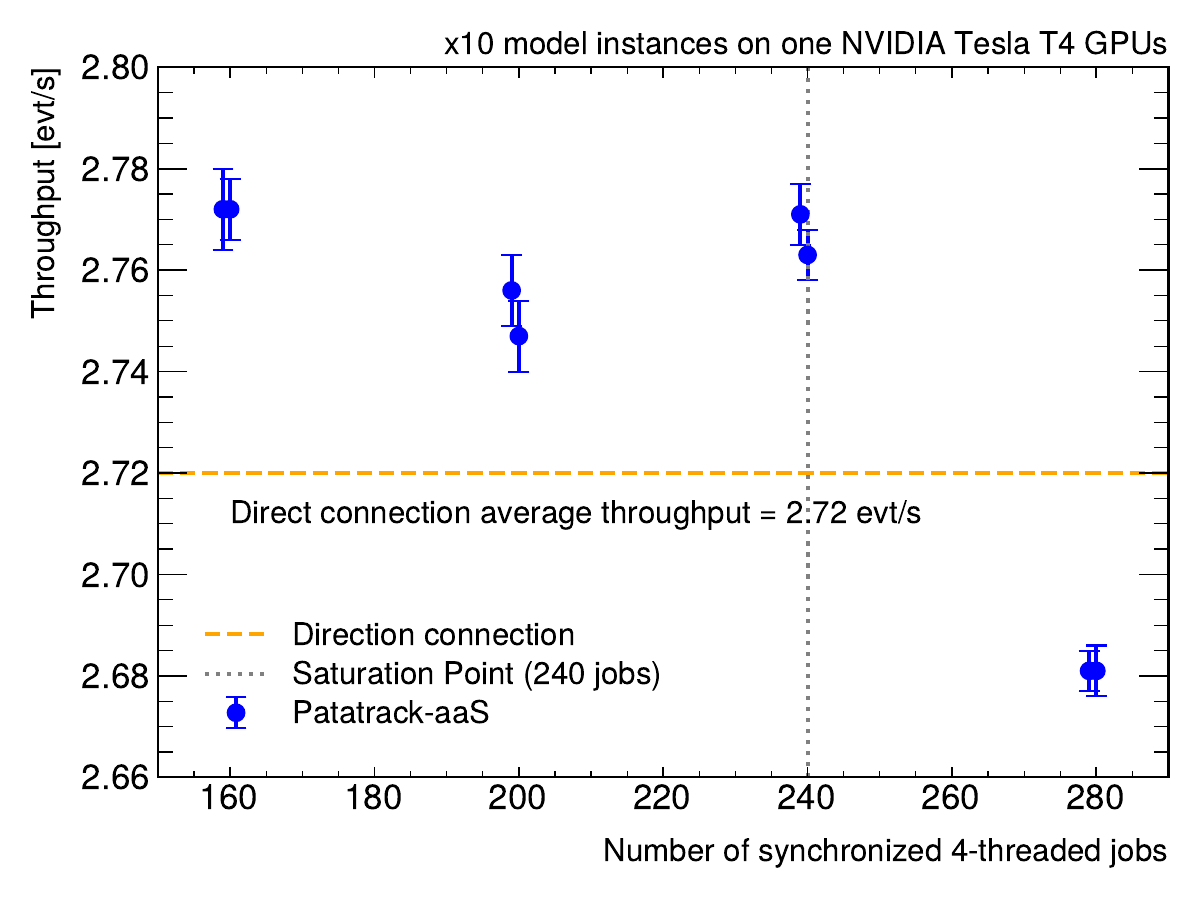}
\caption{A throughput saturation scan is performed by launching a server with ten model instances loaded on one NVIDIA Tesla T4 \ac{GPU}. The remote Triton server loads the Patatrack model, receiving inference requests from multiple synchronized 4-thread CPU client jobs. The throughput is expected to stay at the same level before the \ac{GPU} computing resources are fully saturated. The server becomes fully saturated when around 240 synchronized 4-thread CPU client jobs send requests simultaneously. The throughput starts to drop beyond the saturation point.}
\label{fig:saturation_test}
\end{figure}

Before the \ac{GPU} saturates, the throughput improvement remains flat as the number of CPU clients increases, stabilizing around 10\%. This flat gain is approximately equal to the processing time of the Patatrack tasks offloaded to \ac{GPU}s directly connected to CPUs. Compared to the current \ac{HLT} workflow, which uses 64 CPU cores to serve one \ac{GPU}, the as-a-service approach demonstrates significant potential to serve a larger amount of CPUs simultaneously and increase \ac{GPU} utilization. This improved utilization can reduce the number of \ac{GPU}s needed to serve the same number of clients.

A further test of the throughput saturation scan was conducted with a fully realistic setup of the HLT workflow using an emulated HLT system constructed using Google Cloud. By deploying a server with ten model instances running on a single NVIDIA Tesla T4 \ac{GPU}, the Triton server hosted the Patatrack models and received inference requests from multiple synchronized 4-thread CPU client jobs. Each scan was performed twice per data point to ensure the stability of the results. The throughput remained stable until the \ac{GPU} computing resources reached full saturation. Saturation occurred when 240 synchronized 4-thread CPU client jobs were sent to the server, beyond which the throughput began to decrease. 

Throughput measurements were repeated multiple times at each data point, with the uncertainties calculated as the standard deviation of the measured throughput values for a given test. As shown in Fig~\ref{fig:saturation_test}, the results show that the Patatrack-as-a-Service framework can process up to 240 simultaneous inference requests without experiencing a drop in throughput. A 2\% increase in throughput was observed compared to the average throughput under direct connection, highlighting the server's capability to optimize and handle a higher number of simultaneous inference requests. Moreover, when comparing our approach with the existing CMS HLT GPU model, represented by the orange line, we find that we can more than double the number of threads that a GPU can service.

\section{Exa.TrkX as a Service in ACTS}
\label{sec:exatrkXaas}
The \exatrkX pipeline benefits from the flexibility of a custom backend. Custom backend provides fine-grained control when executing on the NVIDIA Triton Inference Server. This is especially important for sequential pipelines like \exatrkX that connect multiple \ac{ML} models. The client communications with the remote Triton Inference Server are implemented in \ac{ACTS}. 

The standalone implementation includes a configurable class with interfaces for loading the full pipeline, performing inference, and setting the \ac{GPU} device ID. Assigning models and data to a specific \ac{GPU} minimizes data transfers and optimizes performance in multi-\ac{GPU} environments. After this step, the Triton server manages inputs (inference requests) and outputs (responses), streamlining the process for server-side deployment.

\subsection{\exatrkX backend lifecycle}
The \exatrkX backend demonstrates how to use a custom backend as a wrapper around a complicated \ac{ML} workflow, which is common in 
high energy physics. The backend architecture follows a modular life cycle divided into three phases:
\begin{itemize}
    \item \textit{Initialization}: Configurations are loaded, and the model is prepared. During this phase, the server setup begins with the creation of a model object, which is configured according to a predefined input shape, backend library, and model path, as detailed in the configuration. The model instance fetches the device ID from the model object to ensure that all computations are executed on the same \ac{GPU} where the data resides, enhancing processing efficiency.
    
    \item \textit{Execution}: Data is transferred to the \ac{GPU} memory for processing. Model instances are created based on the instance\_group settings in the model configuration. The server opens specific ports to accept inputs, such as a vector of 3D spacepoints position, via HTTP or gRPC protocols. These inputs are processed to generate track candidates.

    \item \textit{Termination}: Resources are cleaned up after processing, and responses are sent back to the client. This phase ensures that all resources are efficiently released and clients receive accurate output from the \exatrkX pipeline.
\end{itemize}

\subsection{Standalone algorithm throughput tests}
A standalone pipeline is used to inspect the performance of the Triton server, including throughput and latency. The tests use the simulated \ttbar events as described in Section~\ref{sec:exatrkxalgo} as input data, and the output track candidates are validated to be identical whether inference occurs on a local \ac{GPU} or on a remote \ac{GPU} accessed via the Triton server. The tests are performed on Perlmutter, a heterogeneous computing system at the National Energy Research Scientific Computing Center (NERSC). For direct inference, the tests are performed on computing nodes directly connected with \ac{GPU}s. For inference using the Triton server, the client is set up on a CPU node while the server is launched on another node with access to a \ac{GPU}. The data is transferred between the client and the server through the gRPC protocol. NVIDIA A100-SXM4 \ac{GPU}s with \qty{40}{\giga\byte} and \qty{80}{\giga\byte} of memory are tested.

\subsubsection{Multiple model instance scaling}
\begin{figure}[ht]
\centering
\includegraphics[width=0.6\textwidth]{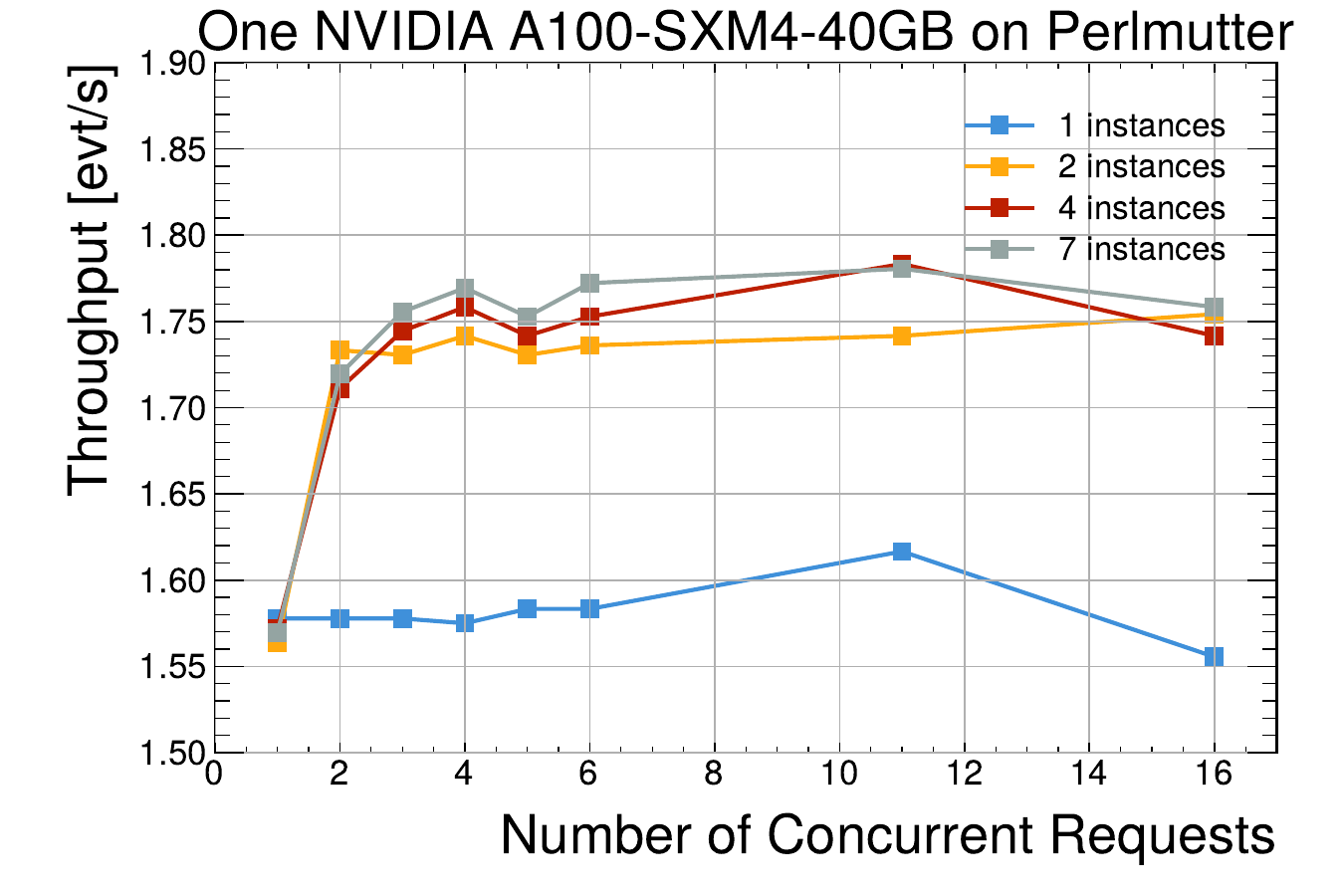}
\caption{Inference throughput with the \exatrkX model on an A100-SXM4-40GB \ac{GPU}. The throughput is measured with a fixed number of Triton instances while increasing the number of concurrent requests until the \ac{GPU} saturates and reaches the maximum throughput. The throughput is compared between different numbers of model instances running on a single \ac{GPU}. It is observed that with two or more model instances on the A100-SXM4 \ac{GPU} all receiving requests, the maximum throughput of about 1.75 events per second can be reached.}
\label{fig:exatrkx_singlegpu_throughput}
\end{figure}

\begin{figure}[ht]
\centering
\includegraphics[width=0.49\textwidth]{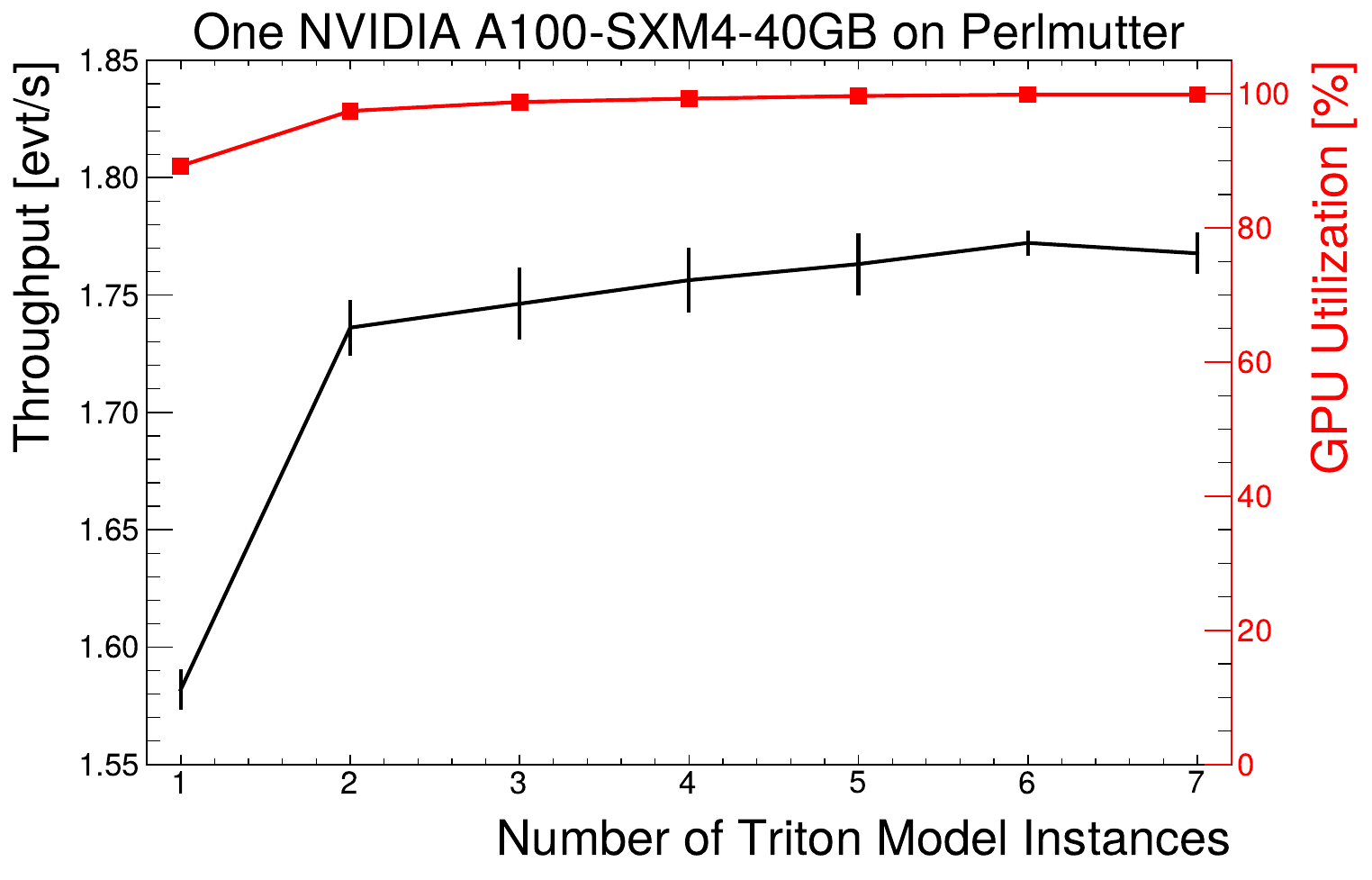}
\includegraphics[width=0.49\textwidth]{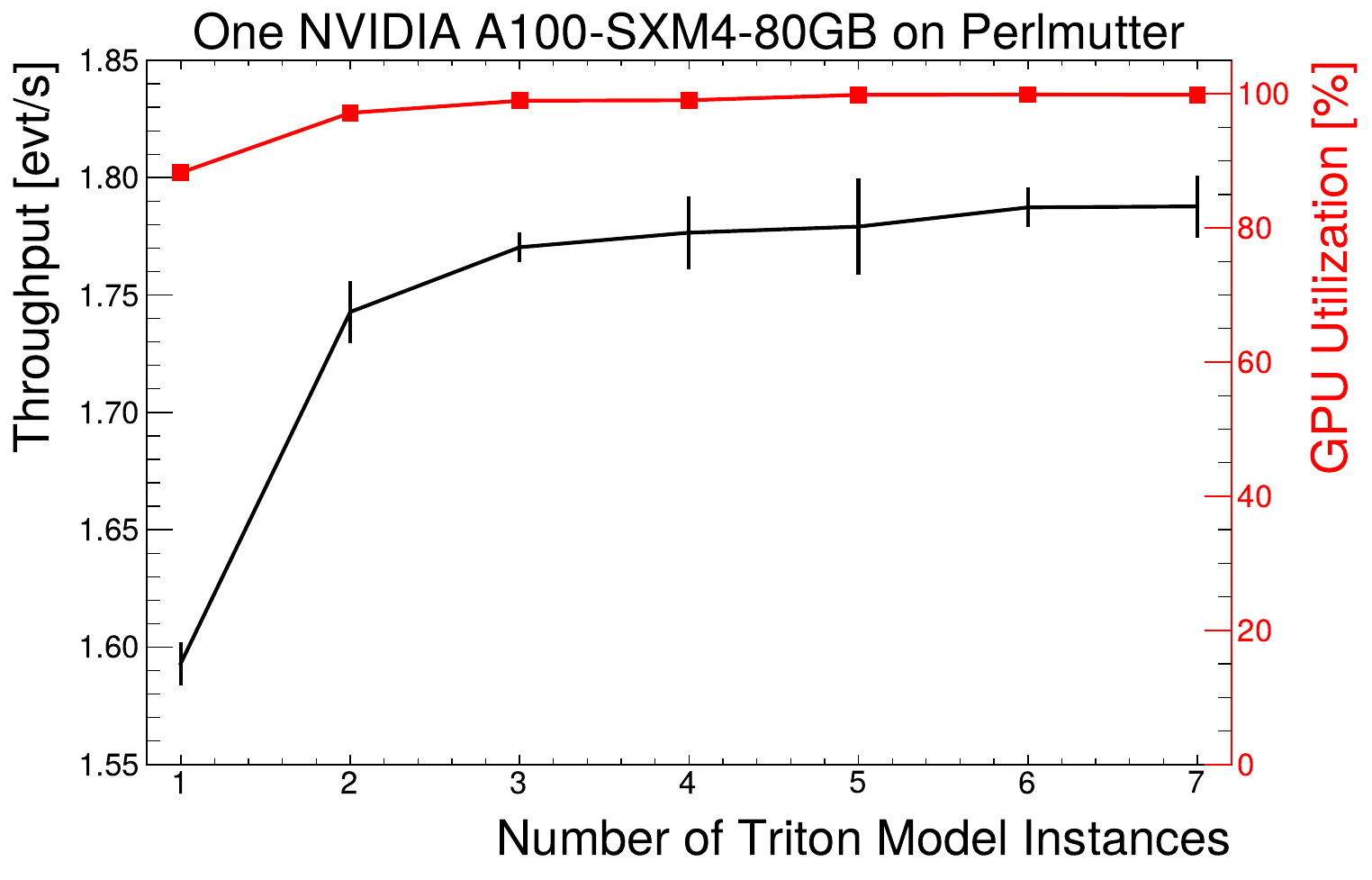}
\caption{Maximum throughput and \ac{GPU} utilization measured on an A100-SXM4-40GB (left) \ac{GPU} and an A100-SXM4-80GB (right) \ac{GPU} with 1 to 7 model instances. No significant throughput difference is observed between the two \ac{GPU}s with different amounts of memory. Each data point represents the average throughput of 10 measurements when the \ac{GPU} reaches the maximum throughput, with the number of concurrent requests ranging from the number of model instances to 10 more concurrent requests beyond the saturation point. The throughput (black line) and \ac{GPU} utilization (red line) reach their maxima simultaneously when there are four or more model instances.}
\label{fig:exatrkx_singlegpu_gpuutil}
\end{figure}

The throughput and \ac{GPU} utilization are measured with the performance analyzer as described in Section~\ref{sec:perfclient}. The client makes asynchronous calls to the server so that the client does not block the thread while waiting for the inference results from the server. The maximum throughput for a single \ac{GPU} is measured to be about 1.75 events per second for both types of \ac{GPU}s, shown in Fig.~\ref{fig:exatrkx_singlegpu_throughput} for the \qty{40}{\giga\byte} \ac{GPU}. Saturation is reached when there are more than 2 model instances on the A100-SXM4 \ac{GPU}s and when the number of concurrent requests is larger than the number of model instances hosted on the Triton server. The throughput increases with the number of model instances because the \ac{GPU} utilization is improved. This point that multiple model instances better utilize the \ac{GPU} is shown in Fig.~\ref{fig:exatrkx_singlegpu_gpuutil}. The latency in the measurements is dominated by computing inference time and queue time. Other components are negligible, including the time that the client sends/receives data, the network sends/receives data, and the server computes input/output. This is expected due to the simple network topology between the client and server in the current setup.

\subsubsection{Multiple GPU scaling}
\begin{figure}[ht]
\centering
\includegraphics[width=0.49\textwidth]{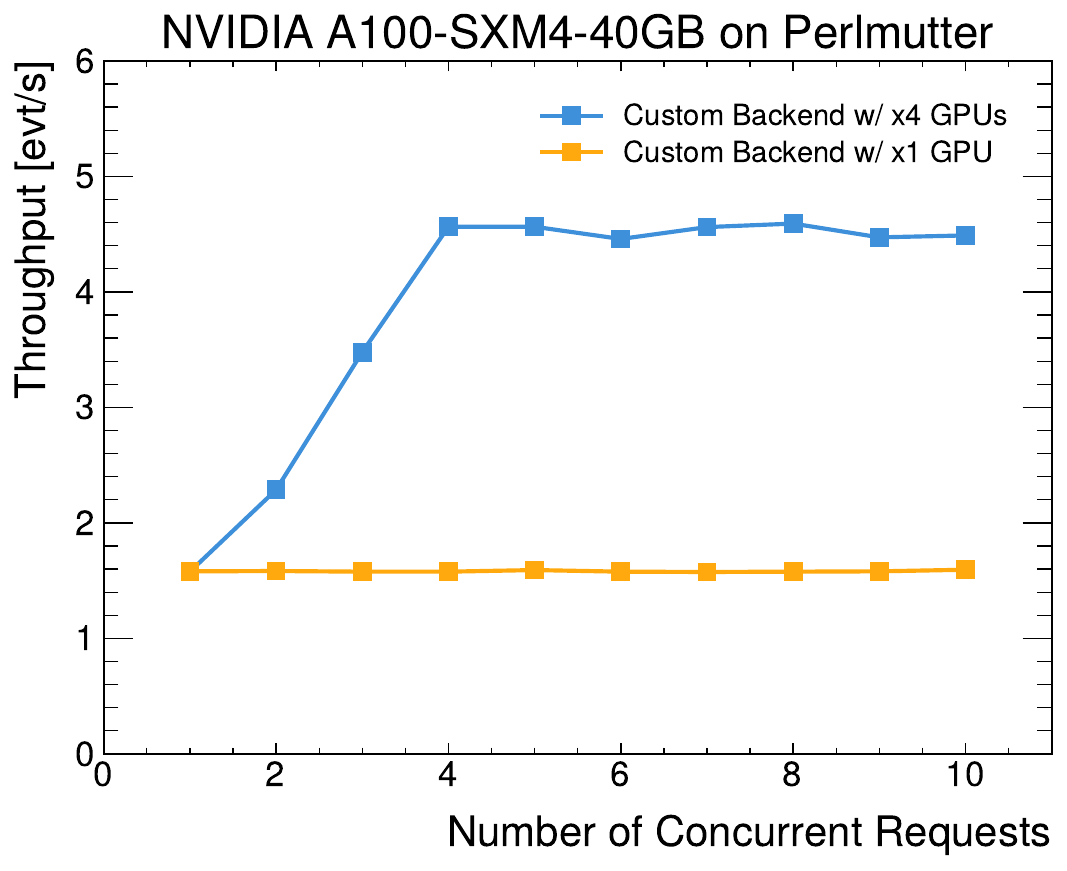}
\caption{Throughput of Triton servers with 1 \ac{GPU} and 4 \ac{GPU}s. The maximum throughput increases from 1.6 events to 4.6 events per second.}
\label{fig:exatrkx_multiple_gpus}
\end{figure}

The throughput for Triton servers with one \ac{GPU} and four \ac{GPU}s are compared. The throughput is measured with one model instance, and all \ac{GPU}s are occupied with requests. We see an increase from 1.6 to 4.6 events per second in Fig.~\ref{fig:exatrkx_multiple_gpus}. The default load balancer in the Triton server is used to distribute requests among all the connected \ac{GPU}s. Further optimization of load balancing is beyond the scope of this study; it may be examined in the future when testing the server deployment at \ac{HPC} centers.  

\subsection{Integrated throughput tests with ACTS}
In the ACTS reconstruction workflow, the \exatrkX pipeline is the dominant component,  consuming a significant fraction of the total processing time. The track reconstruction process can be divided into several steps. The raw measurements from the pixel and strip detectors are converted into three-dimensional space points for pattern recognition, executed in less than \qty{8}{\milli\second} on an AMD EPYC 7763 CPU. 
Subsequently, these space points serve as inputs to the \exatrkX pipeline, which identifies a collection of proto-track candidates. A proto-track collection is simply a list of space points that the \exatrkX pipeline predicts will form a track. The final step of this sequence is the track-fitting process using the Kalman filter.

Three implementations of the \exatrkX pipeline are considered and compared: direct CPU, direct \ac{GPU}, and \ac{GPU} inference as a service. In the first two cases, the \exatrkX pipeline is executed using the local device without interfacing with Triton. It is straightforward to implement an execution sequence that can switch between direct and as-a-service implementations. The mean processing duration was computed from ten simulated \ttbar events, with inference times summarized in Table~\ref{tab:extrkX_time}. Under baseline conditions, the \exatrkX model operates on a dedicated CPU node at Perlmutter, accounting for approximately 95\% of the timeline from raw measurement to track fitting. A significant speedup is achieved using the \ac{GPU} to run inference for \exatrkX pipeline, completed in 2.4 seconds on a directly connected \ac{GPU}. This investigation used an NVIDIA A100 \ac{GPU} for both the direct and as-a-service approaches.

Furthermore, deploying the \exatrkX pipeline on a Triton server introduced no detectable overhead. The throughput is almost identical for inference via direct \ac{GPU} and remote Triton server.
This highlights the robustness of the Triton server implementation for the \exatrkX model and the significant gain in the event processing rate achievable for the \ac{HL-LHC}.

The inference as a service measurement is performed with the client and server located at Perlmutter, so the network latency is expected to be negligible. This is relevant to demonstrate the potential application of this technology in the HLT farm, where the \ac{GPU} cluster is close to the CPU processors, which minimizes the network latency. An alternative scenario could be to utilize a remote \ac{GPU} farm from a powerful \ac{HPC}, such as the National Research Platform at UCSD. In such a scenario, additional latency is expected, which strongly depends on the distance between the client and server that provides the inference calculation.
However, using asynchronous communication between the client and server  minimizes the impact of this latency on the event processing time and throughput, often to a level where it is negligible~\cite{CMS:2024twn}.

\begin{table}[htpb]
    \centering
    \begin{tabular}{c|c}
        Implementation & \exatrkX model inference time (s) \\ \hline
        Direct CPU &  9.65 \\
        Direct \ac{GPU} &  2.42 \\
        \exatrkX-aaS \ac{GPU} &  2.24 \\
    \end{tabular}
    \caption{Average inference time for the \exatrkX model using simulated \ttbar events with an average pileup of 200.}
    \label{tab:extrkX_time}
\end{table}

\section{Summary}
\label{sec:summary}

Track reconstruction is a critical and computationally intensive step in data processing at the \ac{HL-LHC}. 
This paper explores the potential of leveraging modern \ac{GPU}s to address the growing computational challenges through an innovative \emph{inference as a service} approach. Two representative track reconstruction algorithms were evaluated: the ruled-based Patatrack algorithm, utilized for online pixel track and vertex reconstruction, and the \ac{ML}-based \exatrkX pipeline, which aims to be used in online and offline pattern recognition. Both algorithms were successfully adapted to run on the NVIDIA Triton Inference Server, enabling efficient use of \ac{GPU} computing resources and increased throughput. Using custom backend opens up new possibilities, as it not only facilitates the efficient scaling of complex \ac{ML} pipelines with Triton but also extends these capabilities to non-ML algorithms. It enables the seamless scaling of rule-based, non-ML workflows on \acp{GPU}, harnessing their substantial parallel processing potential.

When considering the benefits of using GPU as a service compared to direct GPU or CPU, we note that the GPU-as-a-service paradigm achieves near full GPU utilization; this allows for the maximum observed throughputs to be attained. When comparing directly with a full 64-core CPU, we find an approximate factor of 2 (Patatrack on an NVIDIA T4) to a factor of 4 (Exa.TrkX on an NVIDIA A100) increase in overall throughput. This leads to an approximate 4-8x reduction in power consumption and, ultimately,  operational cost. The equivalent reductions without as-a-service are limited due to the underutilization of the GPU. 

The tracking-as-a-service approach can achieve better throughput with minimal additional overhead compared to the traditional direct-connection approach while significantly improving GPU utilization. This method provides flexibility to dynamically scale resources based on demand, potentially reducing the total number of \acp{GPU} required. The tracking as a service approach enhances the adaptability and sustainability of computing resources, laying the groundwork for more efficient data processing pipelines in the era of the \ac{HL-LHC}. Several future developments are still actively under investigation to bring this portability to other \ac{GPU} vendors and other coprocessor types, ARM processors, \ac{TPU}s, and \ac{FPGA}. There is also ongoing work to optimize the load balancing across multiple \ac{GPU}s and to reduce the overhead further.

\acknowledgments

K. Pedro and N. Tran are supported by Fermi Forward Discovery Group, LLC under Contract No. 89243024CSC000002 with the U.S. Department of Energy, Office of Science, Office of High Energy Physics.
Y. Feng was supported by Fermi Research Alliance, LLC under Contract No. DE-AC02-07CH11359 with the Department of Energy (DOE), Office of Science, Office of High Energy Physics and the DOE Early Career Research Program under Award No. DE-0000247070.
Y. Chou, M. Cochran-Branson, J. Duarte, P. Harris, S. Hsu, M. Liu, P. McCormack, J.-F. Schulte, Y. Yao, and H. Zhao are supported by National Science Foundation (NSF) grant No. PHY-2117997.
M. Liu, P. McCormack, J.-F. Schulte, and Y. Yao were supported by the U.S. CMS Software and Computing Operations Program under the U.S. CMS HL-LHC R\&D Initiative.
Additional support for cloud credits was obtained through the Internet2 Exploring Clouds to accelerate science grant NSF Award \#1904444.
This research used resources of the National Energy Research Scientific Computing Center (NERSC), a U.S. Department of Energy Office of Science User Facility located at Lawrence Berkeley National Laboratory, operated under Contract No. DE-AC02-05CH11231.

%Still missing Purdue

\clearpage
\bibliographystyle{JHEP}
\bibliography{biblio.bib}

\providecommand{\href}[2]{#2}\begingroup\raggedright\begin{thebibliography}{10}

\bibitem{Evans:2008zzb}
L.~Evans and P.~Bryant, \emph{{LHC} machine}, \href{https://doi.org/10.1088/1748-0221/3/08/S08001}{\emph{JINST} {\bfseries 3} (2008) S08001}.

\bibitem{ATLAS:2008xda}
{\scshape ATLAS} collaboration, \emph{The {ATLAS} experiment at the {CERN} {Large Hadron Collider}}, \href{https://doi.org/10.1088/1748-0221/3/08/S08003}{\emph{JINST} {\bfseries 3} (2008) S08003}.

\bibitem{CMS:2008xjf}
{\scshape CMS} collaboration, \emph{The {CMS} experiment at the {CERN LHC}}, \href{https://doi.org/10.1088/1748-0221/3/08/S08004}{\emph{JINST} {\bfseries 3} (2008) S08004}.

\bibitem{Strandlie:2010zz}
A.~Strandlie and R.~Fr{\"{u}}hwirth, \emph{{Track and vertex reconstruction: From classical to adaptive methods}}, \href{https://doi.org/10.1103/RevModPhys.82.1419}{\emph{Rev. Mod. Phys.} {\bfseries 82} (2010) 1419}.

\bibitem{CERN-LHCC-2020-015}
{\scshape ATLAS} collaboration, \emph{{ATLAS HL-LHC Computing Conceptual Design Report}},  Tech. Rep. \href{https://cds.cern.ch/record/2729668}{CERN-LHCC-2020-015}, CERN, Geneva (2020).

\bibitem{Software:2815292}
{CMS Offline Software and Computing}, \emph{{CMS Phase-2 Computing Model: Update Document}},  {CMS Note} \href{https://cds.cern.ch/record/2815292}{CERN-CMS-NOTE-2022-008} (2022).

\bibitem{Duarte:2019fta}
J.~Duarte et~al., \emph{{FPGA-accelerated machine learning inference as a service for particle physics computing}}, \href{https://doi.org/10.1007/s41781-019-0027-2}{\emph{Comput. Softw. Big Sci.} {\bfseries 3} (2019) 13} [\href{https://arxiv.org/abs/1904.08986}{{\ttfamily 1904.08986}}].

\bibitem{Rankin:2020usv}
D.S.~Rankin et~al., \emph{{FPGAs-as-a-Service Toolkit (FaaST)}},  in \emph{{2020 IEEE/ACM International Workshop on Heterogeneous High-performance Reconfigurable Computing (H2RC)}}, 2020, \href{https://doi.org/10.1109/H2RC51942.2020.00010}{DOI} [\href{https://arxiv.org/abs/2010.08556}{{\ttfamily 2010.08556}}].

\bibitem{Krupa:2020bwg}
J.~Krupa et~al., \emph{{GPU coprocessors as a service for deep learning inference in high energy physics}}, \href{https://doi.org/10.1088/2632-2153/abec21}{\emph{Mach. Learn. Sci. Tech.} {\bfseries 2} (2021) 035005} [\href{https://arxiv.org/abs/2007.10359}{{\ttfamily 2007.10359}}].

\bibitem{Savard:2023wwi}
C.~Savard, N.~Manganelli, B.~Holzman, L.~Gray, A.~Perloff, K.~Pedro et~al., \emph{{Optimizing High-Throughput Inference on Graph Neural Networks at Shared Computing Facilities with the NVIDIA Triton Inference Server}}, \href{https://doi.org/10.1007/s41781-024-00123-2}{\emph{Comput. Softw. Big Sci.} {\bfseries 8} (2024) 14} [\href{https://arxiv.org/abs/2312.06838}{{\ttfamily 2312.06838}}].

\bibitem{CMS:2024twn}
{\scshape CMS} collaboration, \emph{{Portable acceleration of CMS computing workflows with coprocessors as a service}}, \href{https://doi.org/10.1007/s41781-024-00124-1}{\emph{Comput. Softw. Big Sci.} {\bfseries 8} (2024) 17} [\href{https://arxiv.org/abs/2402.15366}{{\ttfamily 2402.15366}}].

\bibitem{Wang:2020fjr}
M.~Wang, T.~Yang, M.~Acosta~Flechas, P.~Harris, B.~Hawks, B.~Holzman et~al., \emph{{GPU-Accelerated Machine Learning Inference as a Service for Computing in Neutrino Experiments}}, \href{https://doi.org/10.3389/fdata.2020.604083}{\emph{Front. Big Data} {\bfseries 3} (2021) 604083} [\href{https://arxiv.org/abs/2009.04509}{{\ttfamily 2009.04509}}].

\bibitem{Cai:2023ldc}
T.~Cai, K.~Herner, T.~Yang, M.~Wang, M.A.~Flechas, P.~Harris et~al., \emph{{Accelerating Machine Learning Inference with GPUs in ProtoDUNE Data Processing}}, \href{https://doi.org/10.1007/s41781-023-00101-0}{\emph{Comput. Softw. Big Sci.} {\bfseries 7} (2023) 11} [\href{https://arxiv.org/abs/2301.04633}{{\ttfamily 2301.04633}}].

\bibitem{Gunny:2021gne}
A.~Gunny, D.~Rankin, J.~Krupa, M.~Saleem, T.~Nguyen, M.~Coughlin et~al., \emph{{Hardware-accelerated Inference for Real-Time Gravitational-Wave Astronomy}}, \href{https://doi.org/10.1038/s41550-022-01651-w}{\emph{Nat. Astron.} {\bfseries 6} (2022) 529} [\href{https://arxiv.org/abs/2108.12430}{{\ttfamily 2108.12430}}].

\bibitem{Bocci:2020pmi}
A.~Bocci, V.~Innocente, M.~Kortelainen, F.~Pantaleo and M.~Rovere, \emph{{Heterogeneous Reconstruction of Tracks and Primary Vertices With the CMS Pixel Tracker}}, \href{https://doi.org/10.3389/fdata.2020.601728}{\emph{Front. Big Data} {\bfseries 3} (2020) 601728} [\href{https://arxiv.org/abs/2008.13461}{{\ttfamily 2008.13461}}].

\bibitem{Bocci:2020olh}
A.~Bocci, D.~Dagenhart, V.~Innocente, C.~Jones, M.~Kortelainen, F.~Pantaleo et~al., \emph{{Bringing heterogeneity to the CMS software framework}}, \href{https://doi.org/10.1051/epjconf/202024505009}{\emph{EPJ Web Conf.} {\bfseries 245} (2020) 05009} [\href{https://arxiv.org/abs/2004.04334}{{\ttfamily 2004.04334}}].

\bibitem{ExaTrkX:2021abe}
{\scshape Exa.TrkX} collaboration, \emph{{Performance of a geometric deep learning pipeline for HL-LHC particle tracking}}, \href{https://doi.org/10.1140/epjc/s10052-021-09675-8}{\emph{Eur. Phys. J. C} {\bfseries 81} (2021) 876} [\href{https://arxiv.org/abs/2103.06995}{{\ttfamily 2103.06995}}].

\bibitem{ATLAS:2020esi}
{\scshape ATLAS} collaboration, \emph{{Operation of the ATLAS trigger system in Run 2}}, \href{https://doi.org/10.1088/1748-0221/15/10/P10004}{\emph{JINST} {\bfseries 15} (2020) P10004} [\href{https://arxiv.org/abs/2007.12539}{{\ttfamily 2007.12539}}].

\bibitem{CMS:2020cmk}
{\scshape CMS} collaboration, \emph{{Performance of the CMS Level-1 trigger in proton-proton collisions at $\sqrt{s} =$ 13 TeV}}, \href{https://doi.org/10.1088/1748-0221/15/10/P10017}{\emph{JINST} {\bfseries 15} (2020) P10017} [\href{https://arxiv.org/abs/2006.10165}{{\ttfamily 2006.10165}}].

\bibitem{CMS:2024aqx}
{\scshape CMS} collaboration, \emph{{Performance of the CMS high-level trigger during LHC Run 2}}, \href{https://doi.org/10.1088/1748-0221/19/11/P11021}{\emph{JINST} {\bfseries 19} (2024) P11021} [\href{https://arxiv.org/abs/2410.17038}{{\ttfamily 2410.17038}}].

\bibitem{Collaboration:2759072}
{\scshape CMS} collaboration, \emph{{The Phase-2 Upgrade of the CMS Data Acquisition and High Level Trigger}},  CMS Technical Design Report \href{https://cds.cern.ch/record/2759072}{{CERN-LHCC-2021-007, CMS-TDR-022}} (2021).

\bibitem{Zabi:2020gjd}
{\scshape CMS} collaboration, \emph{{The Phase-2 Upgrade of the CMS Level-1 Trigger}},  CMS Technical Design Report \href{https://cds.cern.ch/record/2714892}{CERN-LHCC-2020-004, CMS-TDR-021} (2020).

\bibitem{Bronstein_2017}
M.M.~Bronstein, J.~Bruna, Y.~LeCun, A.~Szlam and P.~Vandergheynst, \emph{Geometric deep learning: Going beyond {Euclidean} data}, \href{https://doi.org/10.1109/msp.2017.2693418}{\emph{IEEE Signal Processing Magazine} {\bfseries 34} (2017) 18}.

\bibitem{Dezoort:2021kfk}
G.~Dezoort, S.~Thais, I.~Ojalvo, P.~Elmer, V.~Razavimaleki, J.~Duarte et~al., \emph{Charged particle tracking via edge-classifying interaction networks}, \href{https://doi.org/10.1007/s41781-021-00073-z}{\emph{Comput. Softw. Big Sci.} {\bfseries 5} (2021) 26} [\href{https://arxiv.org/abs/2103.16701}{{\ttfamily 2103.16701}}].

\bibitem{Lieret:2023aqg}
K.~Lieret, G.~DeZoort, D.~Chatterjee, J.~Park, S.~Miao and P.~Li, \emph{{High Pileup Particle Tracking with Object Condensation}},  in \emph{{8th International Connecting The Dots Workshop}}, 2023 [\href{https://arxiv.org/abs/2312.03823}{{\ttfamily 2312.03823}}].

\bibitem{Liu:2023siw}
R.~Liu, P.~Calafiura, S.~Farrell, X.~Ju, D.T.~Murnane and T.M.~Pham, \emph{{Hierarchical Graph Neural Networks for Particle Track Reconstruction}},  in \emph{{21st International Workshop on Advanced Computing and Analysis Techniques in Physics Research}}, 2023 [\href{https://arxiv.org/abs/2303.01640}{{\ttfamily 2303.01640}}].

\bibitem{miao2024localitysensitive}
S.~Miao, Z.~Lu, M.~Liu, J.~Duarte and P.~Li, \emph{{Locality-Sensitive Hashing-Based Efficient Point Transformer with Applications in High-Energy Physics}},  in \emph{{41st International Conference on Machine Learning}}, vol.~235, p.~35546, 2024, \href{https://proceedings.mlr.press/v235/miao24b.html}{https://proceedings.mlr.press/v235/miao24b.html} [\href{https://arxiv.org/abs/2402.12535}{{\ttfamily 2402.12535}}].

\bibitem{gnn:IN}
P.W.~Battaglia, J.B.~Hamrick, V.~Bapst, A.~Sanchez-Gonzalez, V.~Zambaldi, M.~Malinowski et~al., \emph{Relational inductive biases, deep learning, and graph networks},  \href{https://arxiv.org/abs/1806.01261}{{\ttfamily 1806.01261}}.

\bibitem{frnn}
L.~Xue, ``{FRNN}.'' \url{https://github.com/lxxue/FRNN}, 2013.

\bibitem{douze2024faiss}
M.~Douze, A.~Guzhva, C.~Deng, J.~Johnson, G.~Szilvasy, P.-E.~Mazaré et~al., \emph{The faiss library},  \href{https://arxiv.org/abs/2401.08281}{{\ttfamily 2401.08281}}.

\bibitem{cugraph}
cuGRAPH, ``{RAPIDS Graph documentation}.'' \url{https://docs.rapids.ai/api/cugraph/stable/}.

\bibitem{Ai:2021ghi}
X.~Ai et~al., \emph{{A Common Tracking Software Project}}, \href{https://doi.org/10.1007/s41781-021-00078-8}{\emph{Comput. Softw. Big Sci.} {\bfseries 6} (2022) 8} [\href{https://arxiv.org/abs/2106.13593}{{\ttfamily 2106.13593}}].

\bibitem{Edmonds:2008zz}
K.~Edmonds, S.~Fleischmann, T.~Lenz, C.~Magass, J.~Mechnich and A.~Salzburger, \emph{{The fast ATLAS track simulation (FATRAS)}}, .

\bibitem{Bierlich:2022pfr}
C.~Bierlich et~al., \emph{{A comprehensive guide to the physics and usage of PYTHIA 8.3}}, \href{https://doi.org/10.21468/SciPostPhysCodeb.8}{\emph{SciPost Phys. Codeb.} {\bfseries 2022} (2022) 8} [\href{https://arxiv.org/abs/2203.11601}{{\ttfamily 2203.11601}}].

\bibitem{triton}
NVIDIA, ``{NVIDIA Triton Inference Server}.'' \url{https://docs.nvidia.com/deeplearning/triton-inference-server/user-guide/docs/index.html}.

\bibitem{grpc}
Google, ``{gRPC: A high performance, open source universal RPC framework}.'' \url{https://grpc.io/}.

\bibitem{pytorch}
A.~Paszke et~al., \emph{{PyTorch}: An imperative style, high-performance deep learning library},  in \emph{Advances in Neural Information Processing Systems 32}, H.~Wallach, H.~Larochelle, A.~Beygelzimer, F.~d\textquotesingle Alch\'{e}-Buc, E.~Fox and R.~Garnett, eds., p.~8024, Curran Associates, Inc., 2019, \href{http://papers.neurips.cc/paper/9015-pytorch-an-imperative-style-high-performance-deep-learning-library.pdf}{http://papers.neurips.cc/paper/9015-pytorch-an-imperative-style-high-performance-deep-learning-library.pdf} [\href{https://arxiv.org/abs/1912.01703}{{\ttfamily 1912.01703}}].

\bibitem{tensorflow}
M.~Abadi et~al., ``{TensorFlow: A system for large-scale machine learning}.'' 2016.

\bibitem{tensorrt}
NVIDIA, ``{NVIDIA TensorRT}.'' \url{https://developer.nvidia.com/tensorrt}.

\bibitem{onnx}
ONNX, ``{Open Neural Network Exchange (ONNX)}.'' \url{https://github.com/onnx/onnx}.

\bibitem{Zhao:2024uwh}
H.~Zhao et~al., \emph{{Graph Neural Network-based Tracking as a Service}},  in \emph{{Connecting The Dots 2023}}, 2, 2024 [\href{https://arxiv.org/abs/2402.09633}{{\ttfamily 2402.09633}}].

\end{thebibliography}\endgroup

\end{document}